\documentclass[12pt,ngerman,english]{article}
\usepackage[T1]{fontenc}
\usepackage[latin9]{inputenc}
\usepackage[a4paper]{geometry}
\usepackage{color}
\usepackage{babel}
\usepackage{float}
\usepackage{textcomp}
\usepackage{amsmath}
\usepackage{amssymb}
\usepackage{graphicx}
\usepackage[unicode=true,
 bookmarks=true,bookmarksnumbered=true,bookmarksopen=true,bookmarksopenlevel=1,
 breaklinks=false,pdfborder={0 0 1},backref=false,colorlinks=true]
 {hyperref}
\hypersetup{pdftitle={XCA},
 pdfauthor={W.Eckel},
 pdfsubject={Druckst�rke},
 linkcolor=black,citecolor=black,urlcolor=blue,filecolor=blue,pdfpagelayout=OneColumn,pdfnewwindow=true,pdfstartview=XYZ,plainpages=false}
\usepackage{breakurl}

\makeatletter

\newcommand{\noun}[1]{\textsc{#1}}
\newcommand{\lyxmathsym}[1]{\ifmmode\begingroup\def\b@ld{bold}
  \text{\ifx\math@version\b@ld\bfseries\fi#1}\endgroup\else#1\fi}

\newcommand{\lyxaddress}[1]{
\par {\raggedright #1
\vspace{1.4em}
\noindent\par}
}
\newenvironment{lyxcode}
{\par\begin{list}{}{
\setlength{\rightmargin}{\leftmargin}
\setlength{\listparindent}{0pt}
\raggedright
\setlength{\itemsep}{0pt}
\setlength{\parsep}{0pt}
\normalfont\ttfamily}%
 \item[]}
{\end{list}}

\@ifundefined{date}{}{\date{}}
\usepackage{babel}
\usepackage{ifpdf}
\ifpdf 

 \IfFileExists{lmodern.sty}{\usepackage{lmodern}}{}

 \fi 

\let\myTOC\tableofcontents
\renewcommand{\tableofcontents}{%
  \pdfbookmark[1]{\contentsname}{}
  \myTOC
  \cleardoublepage
  \pagenumbering{arabic} }

\makeatother

\usepackage{listings}
\addto\captionsenglish{}
\addto\captionsngerman{}

\begin{document}

\title{Creating an Artificial World with a New Kind of Cellular Automaton}

\author{Walter Eckel }

\maketitle

\lyxaddress{Weil im Schoenbuch}
\begin{abstract}
This paper describes a new concept of cellular automaton (CA). XCA
consists of a set of arcs (edges) that correspond to cells in CA.
At a particular time, the arcs are connected to a directed graph.
With each time step, the arcs exchange their neighbors (adjacent arcs)
according to rules that depend on the statuses of the adjacent arcs.

An XCA can be used to simulate an artificial world beginning with
a Big Bang. In contrast to an CA, an XCA does not require a grid.
However, it can create one, just as the real universe after the Big
Bang generated its own space, which had not excisted previously. Examples
using different rules reveal the manifold nature of the XCA concept.
Similar to John Conway\textquoteright s well-known The Game of Life
simulates birth, survival, and death, this game can simulate a system
that begins from a singularity, and evolves into a complex space.


\newpage{}\tableofcontents{} 
\end{abstract}

\section{Introduction }

This paper describes a new concept of cellular automaton (CA), the
extended CA (XCA), which leads to two paradigm shifts: 1) a more generalized
definition of cell, and 2) different roles for the nodes and arcs
of a network. 
\begin{description}
\item [{Definition\nobreakdash-of\nobreakdash-Cell:}] In general, a cell
is understood as the face of a two-dimensional grid, or more generally,
as the n-dimensional solid of an n-dimensional grid. An XCA requires
a more generalized definition of a cell as follows: 1) a cell is surrounded
by borders 2) any two neighboring cells have a common border. - With
this definition. not only the fields of a grid, but also the nodes
and arcs (edges) of a graph can be considered to be cells. 
\item [{Nodes\nobreakdash-and\nobreakdash-Arcs:}] Normally, the nodes
of a graph represent objects, and the arcs are seen as connections
between objects. With our XCA view, in contrast to the standard view,
the arcs are seen as objects, which are connected at nodes. With this
concept, the number of arcs remains constant, while the number of
nodes may change with each step of evolution. 
\end{description}
In addition to the classical concept of CA, there are other concepts
of CA in which nodes are considered cells. Two neighboring nodes have
an arc as a common border. In contrast to this concept for the arc-based
XCA, two neighboring arcs have a node as a common border.

For each time step, the value of each cell changes according to the
values of the neighboring cells and according to some transition rules.
This is true for both concepts. However, in contrast to a CA, an XCA
can exchange the neighbors (adjacent arcs) according to rules that
depend on the statuses of adjacent arcs.

\begin{figure}[th]
\ \ \ \ \ \ \ \ \ \ \includegraphics{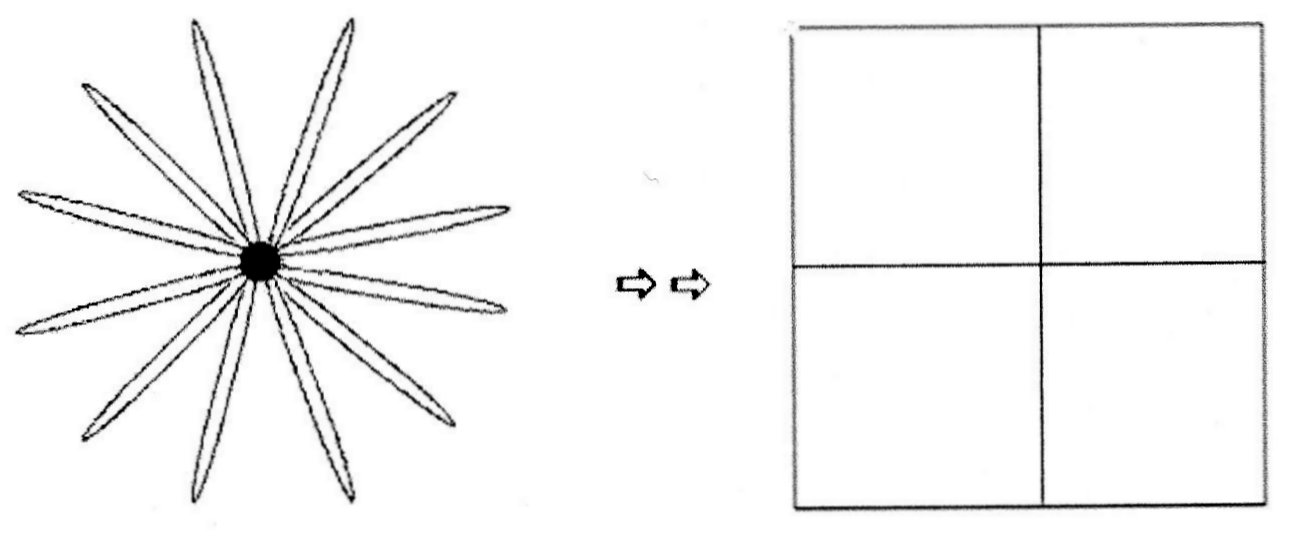}

\caption{\label{fig:Big-Bang-Model-1}Big Bang Model}
\end{figure}

The new model might help to explain physical phenomena. At the beginning
of the universe, all arcs are connected with both ends to a single
node. With the next discrete time steps, the graph alters its structure,
while the arcs maintain their identities (Figure \ref{fig:Big-Bang-Model-1}).
According to this hypothesis, all phenomena are represented by different
configurations and evolutions of arcs. However, not only matter, particles,
and interactions, but also space itself, are represented by arcs.

\section{CA}

In 1970 John H. Conway published \textquotedblleft The Game of Life\textquotedblright{}
\cite{key-7}, which soon gained much popularity. This \textquotedbl{}game\textquotedbl{}
is the best-known example of a CA\cite{key-8}. \textquotedblleft The
Game of Life (or simply Life) is not a game in the conventional sense.
There are no players, and no winning or losing. Once the \textquoteleft pieces\textquoteright{}
are placed in the starting position, the rules determine everything
that happens later\textquotedblright{} \cite{key-9}. The \textquoteleft game\textquoteright{}
is a zero-player game, meaning that its evolution is determined by
its initial state, requiring no further input from humans. One interacts
with The Game of Life by creating an initial configuration and observing
how it evolves.\textquotedblright{} \cite{key-10}

Life is an example of self-organization \cite{key-11}.

A CA consists of an n-dimensional grid, whose cells are assigned specific
values at particular times. For each time step, the value of each
cell changes according to the values of the neighboring cells (usually
including the cell itself) and according to some transition rules.
Different examples of a CA show how chaos can evolve into a structured
organization.

For some examples, visit \href{http://www.collidoscope.com/modernca/}{http://www.collidoscope.com/modernca/}
\cite{key-12}. You can find an introduction to the general concept
of CA in Appendix A.1.

Inspired by the idea of \emph{CA,} I developed an \noun{XCA} with
which an artificial world can be simulated beginning with a \emph{Big
Bang}. In contrast to a CA, an XCA does not require a grid. However,
it can create one, just as the real universe after the Big Bang generated
its own space, which had not existed previously.

\section{XCA}

The arcs of an XCA correspond to the cells in CA. At a particular
time, the arcs are connected to a directed graph. Arcs that are connected
to a node are neighbors. Just as values are assigned to cells of a
CA, particular values can be assigned to the endpoints of an arc.
With each time step, the arcs exchange their neighbors (adjacent arcs)
according to rules that depend on the statuses of adjacent arcs.

Let us denote the start point (head) and end point (tail) of arc $A$
by $A_{1}$ and $A_{0}$, respectively. The values $V(A_{0})$ and
$V(A_{1})$ can be assigned to the start and end points, respectively,
and $V(A)$ to the arc itself. At a particular time, two arcs $A$
and $B$ meet at a node with their start and end points $A_{1}$ and
$B_{0}$, respectively. The state of the node is then denoted as $\left\{ A_{1},\,B_{0}\right\} $,
or alternatively, as $A_{1}B_{0}$.

An XCA game is described using four types of rules: 
\begin{itemize}
\item Cellular rules 
\item Transition rules 
\item Decision and constraint rules 
\item Connectivity 
\end{itemize}

\subsection{Cellular Rules }

Cellular rules apply to the values $V(x)$ and correspond to the transition
rules applied to the cell values of a conventional CA. Consider the
dual graph of a rectangular grid. XCA and CA are then equivalent in
the following conditions: 
\begin{itemize}
\item The grid does not change its topology in discrete time steps. 
\item The nodes in the dual graph correspond to cells, and arcs designate
neighbors of the CA. 
\item If the arcs $a$, $b$, $c$, and $d$ meet at node \{$a_{1},b_{0},c_{0},d_{1}\}$,
then $V(a_{1})=V(b_{0})=V(c_{0})=V(d_{1})$. 
\end{itemize}

\subsection{Transition Rules }

Two arcs can be linked to a node in four possible manners: $A_{1}B_{0}$,
$A_{0}B_{1}$, $A_{1}B_{1}$, and $A_{0}B_{0}$.

\begin{figure}[H]
\noindent \begin{centering}
\includegraphics{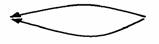} 
\par\end{centering}

\caption{\label{fig:fig1}}
\end{figure}

\begin{figure}[H]
\noindent \begin{centering}
\includegraphics{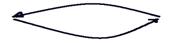} 
\par\end{centering}

\caption{\label{fig:fig2}}
\end{figure}

For $A_{1}B_{1}$, for example, there are four possible transition
rules regarding how the links can change with the next time step: 
\begin{lyxcode}
\begin{minipage}[t]{1\columnwidth}%
\begin{lyxcode}
$A_{1}B_{1}->A_{1}B_{1}$(no~change)~\foreignlanguage{ngerman}{$\Rightarrow$}~-{}-{}-{}-{}-$A$-{}-{}->.<-{}-{}-$B$-{}-{}-{}-{}-~

$A_{1}B_{1}->A_{1}B_{0}$~\foreignlanguage{ngerman}{$\Rightarrow$}~-{}-{}-{}-{}-$A$-{}-{}->.-{}-{}-{}-{}-$B$-{}-{}-{}->~

$A_{1}B_{1}->A_{0}B_{1}$~\foreignlanguage{ngerman}{$\Rightarrow$}~<-{}-{}-{}-$A$-{}-{}-{}-.<-{}-{}-{}-$B$-{}-{}-{}-{}-~

$A_{1}B_{1}->A_{0}B_{0}$~\foreignlanguage{ngerman}{$\Rightarrow$}~<-{}-{}-{}-$A$-{}-{}-{}-.-{}-{}-{}-{}-$B$-{}-{}-{}->~\end{lyxcode}
\end{minipage}
\end{lyxcode}
The two arcs can also be linked together at both sides (multiple arcs)
at the same time, for instance, at $A_{1}B_{1}$ and $A_{0}B_{0}$
to produce a state denoted as $A_{1}B_{1}/A_{0}B_{0}$. Figures \ref{fig:fig1}
and \ref{fig:fig2} show the states $A_{1}B_{1}/A_{0}B_{0}$ and $A_{1}B_{0}/A_{0}B_{1}$,
respectively.

An arc can also be linked to itself with node $\left\{ A_{1}A_{0}\right\} $
by building a circle (loop). The end points $A_{1}$ and $A_{0}$
can connect and disconnect from step to step. If the loop of $A$
is linked to the arc $B$, then the following transition is possible:\\
 $\left\{ A_{i},\,A_{k},\,B_{i}\right\} \Rightarrow\left\{ A_{i},\,B_{i}\right\} /\left\{ A_{k}\right\} $
(the loop opens)\\

In general, the following transition rules are possible for state
$A_{i}B_{k}$, where $i$ and $k$ are either zero or one, and $i$
is not $k$: \\

$A_{i}B_{k\,}$\foreignlanguage{ngerman}{$\Rightarrow$} $A_{i}B_{k}$
(no change)

$A_{i}B_{k\,}$\foreignlanguage{ngerman}{$\Rightarrow$} $A_{i}B_{i}$

$A_{i}B_{k}$ \foreignlanguage{ngerman}{$\Rightarrow$} $A_{k}B_{k}$

$A_{i}B_{i}$\foreignlanguage{ngerman}{ $\Rightarrow$} $A_{i}B_{i}$
(no change)

$A_{i}B_{i}$ \foreignlanguage{ngerman}{$\Rightarrow$} $A_{i}B_{k}$

$A_{i}B_{i}$ \foreignlanguage{ngerman}{$\Rightarrow$} $A_{k}B_{i}$

\subsection{Decision and Constraint Rules}

Consider the transition rule $A_{1}B_{1}$\foreignlanguage{ngerman}{$\Rightarrow$}$A_{1}B_{0}$
being applied to the arcs $a$, $b$, and $c$, meeting at point ${a_{1},b_{1},c_{1}}$.
Arc $a$ can interact with either $b$ or $c$, indvidually or with
both $b$ and $c$ at the same time. If $a$ interacts with $b$,
then there are two possible manners in which the arcs can interact
with the next time step:

$a_{1}b_{1}$\foreignlanguage{ngerman}{ $\Rightarrow$} $a_{1}b_{0}$

$b_{1}a1$\foreignlanguage{ngerman}{ $\Rightarrow$} $b_{1}a_{0}$

This example above shows why decision rules are necessary.\\

\paragraph{Examples of Decision Rules: }
\begin{enumerate}
\item The arcs interact by chance with preferences. 
\item The transition is controlled by the statuses of neighboring arcs. 
\item The status of time step t must not be equal to that of time step t
+2. Otherwise, the arcs interact by chance. 
\item Each arc is identified by a number. At even time step t, the lower
numbered arc has priority. At odd time step t +1, the higher numbered
arc has priority. 
\item The transition is controlled by cellular rules. 
\item Consider the arcs $a$, $b$, $c$, and $d$ are connected at the
nodes $d_{1}c_{0}$, $c_{1}b_{0}$, $b_{1}a_{0}$. By the transition
rules $d_{1}c_{0}$ \foreignlanguage{ngerman}{$\Rightarrow$} $d_{1}c_{1}$,
$c_{1}b_{0}$ \foreignlanguage{ngerman}{$\Rightarrow$} $c_{1}b_{1}$,
and $b_{1}a_{0}$ \foreignlanguage{ngerman}{$\Rightarrow$} $b_{1}a_{1}$
the arcs will meet at node ${d_{1},c_{1},b_{1},a_{1}}$. Now the arc
$c$ becomes a neighbor of $a$. The fact that $d$ is a new neighbor
of $a$ can be understood as $d$ being ranked lower than $b$ and
$c$. This ranking can also be a criterion for a decision rule. 
\end{enumerate}

\paragraph{Examples of Constraints: }
\begin{enumerate}
\item Transitions cannot allow multiple arcs (two arcs being connected at
both ends, see Figures \ref{fig:fig1} and \ref{fig:fig2} above). 
\item Transitions cannot allow circles (self loops). 
\item Transitions cannot allow reversing, meaning that the status of two
arcs at time step t +1 cannot be the same as at time step t -1. 
\item Transitions cannot allow multiple interactions, meaning that an arc
cannot interact with more than one other arc at the same time step. 
\item Any two linked arcs must change their status at every time. 
\end{enumerate}

\subsection{Connectivity }

An XCA graph must not be disconnected at any time step. For a connected
graph, this condition is always true as long as no transition rule
causes a disconnection. If a node $A_{i}B_{k}$ exists, then no transition
is allowed that disconnects A and B, with one exception:

Let us assume that the three arcs $A$, $B$, and $C$ are connected
together with one of their end points. The arc $A$ can then move
to the other end of $B$, thus disconnecting $A$ from $C$. In this
case, the transition $AC\Rightarrow0$ is allowed.

\section{Turing Machine}

A Turing machine models the operation of a computer in a particularly
simple and mathematically analyzable manner.

The machine consists of a tape with an endless sequence of symbols.
At each time step, a tape head reads a single symbol on the tape.
Depending on this symbol, the status of the tape head, and a set of
instructions, the tape head overwrites the symbol with a new one.
Then, it moves to the next symbol either to the left or to the right.

An XCA configuration can emulate a Turing machine. A chain of arcs
whose start points are all in the same direction, represents the tape.
The arcs represent symbols of the tape and do not interact with each
other. A different arc behaves in a manner similar to that of the
tape head that meets the start point of an arc of the chain. Depending
on the value of the symbol arc, the status of the tape head arc, and
a set of instructions, the value of the symbol arc is replaced by
a new one. Then the tape head arc moves either to the left or to the
right of the symbol arc.

In general, any three arcs of a node in an XCA system can be seen
as a type of Turing machine. Two of the arcs correspond to neighboring
symbols of the machine. The third arc corresponds to the tape head.
This type of machine is valid for only a limited time. After this
time, the roles of the arcs can be interchanged. One of the former
symbol arcs becomes the tape head arc and the previous tape head arc
becomes a symbol arc.

Finally, each arc of an XCA can resemble a computer, processing its
neighbors.

\section{XCA Rule Examples}

I like to believe, that XCAs will make it possible to simulate the
physical world, provided that appropriate rules are found. This is
not yet the case, but the following examples will show how multifold
this system is.

\subsection{Rule Example 1}

The arcs interact by chance.

There are no constraints.

The graph must not be disconnected at any time step.

\paragraph{Start Configuration: \protect \protect \protect \protect \protect
\protect \\
 }

At the beginning, all arcs are connected with both ends to a single
node.

\begin{figure}[th]
\ \ \ \ \ \ \ \ \ \ \ \ \ \ \ \ \ \ \ \ \ \ \ \ \ \ \includegraphics[scale=0.6]{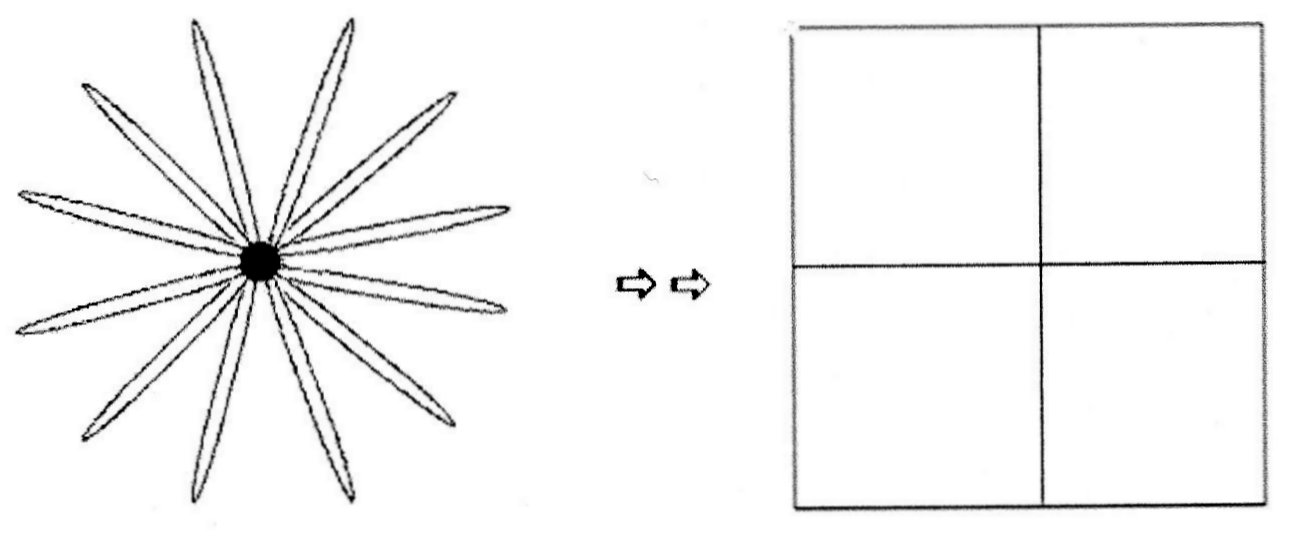}

\caption{\label{fig:Big-Bang-Model}Big Bang Model}
\end{figure}

Because any arc with both ends connected to a single node is adjacent
to any other arc, the arbitrary transitions can result in any possible
configuration. In Figure \ref{fig:Big-Bang-Model}, the initial graph
with 12 arcs evolves into a rectangular grid with 12 arcs by chance.
With the same probability, any other configuration is possible, provided
that the number of arcs remains constant, and the graph is not disconnected.

\subsection{Rule Example 2}

Consider the following transition rules:

$A_{1}B_{1}$ \foreignlanguage{ngerman}{$\Rightarrow$} $A_{0}B_{1}$;

$A_{0}B_{1}$ \foreignlanguage{ngerman}{$\Rightarrow$} $A_{0}B_{0}$;

$A_{0}B_{0}$ \foreignlanguage{ngerman}{$\Rightarrow$} $A_{1}B_{0}$;

$A_{1}B_{0}$ \foreignlanguage{ngerman}{$\Rightarrow$} $A_{1}B_{1}$;
\\
 \\
 As constraints, the transitions do not allow the following: 
\begin{enumerate}
\item multiple arcs (two arcs being connected at both ends, see Figures\ref{fig:fig1}
and \ref{fig:fig2}above) 
\item circles 
\item reversing (the status of two arcs at time step t +1 being the same
as at time step t -1) 
\item multiple interactions (an arc interacting with more than one other
arc at the same time step) 
\end{enumerate}
We can apply these rules to different configurations as follows.

\subsubsection{Configuration 1: }

\begin{figure}[h]
\noindent \begin{centering}
\includegraphics{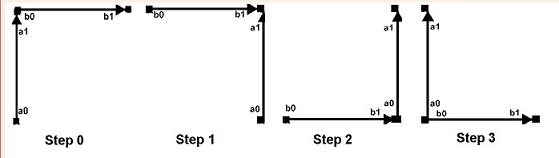} 
\par\end{centering}

\caption{\label{fig:fig3-1}}
\end{figure}

The two arcs $a$ and $b$ meet at point $a_{1}b_{0}$ at the first
time step. Figure \ref{fig:fig3-1} shows steps~0 through 3 (step~4
is equal to step~0).

\noindent \qquad{}

\begin{figure}[H]
\noindent \begin{centering}
\includegraphics[scale=0.8]{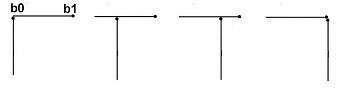} 
\par\end{centering}

\caption{\label{fig:fig4}}
\end{figure}

The following figures show how the graphs of consecutive time steps
evolve through intermediate steps being inserted. In Figure \ref{fig:fig4},
for instance, two intermediate steps are inserted. The two arcs $a$
and $b$ meet at point $a_{1}b_{0}$ at the first time step. Figure
\ref{fig:fig4} shows how the start point of arc $a$ (vertical) moves
from the end point $b_{0}$ to the start point $b_{1}$, passing the
two intermediate points between $b_{0}$ and $b_{1}$.

\qquad{}

\subsubsection{Configuration 2: }

\begin{figure}[H]
\noindent \begin{centering}
\includegraphics{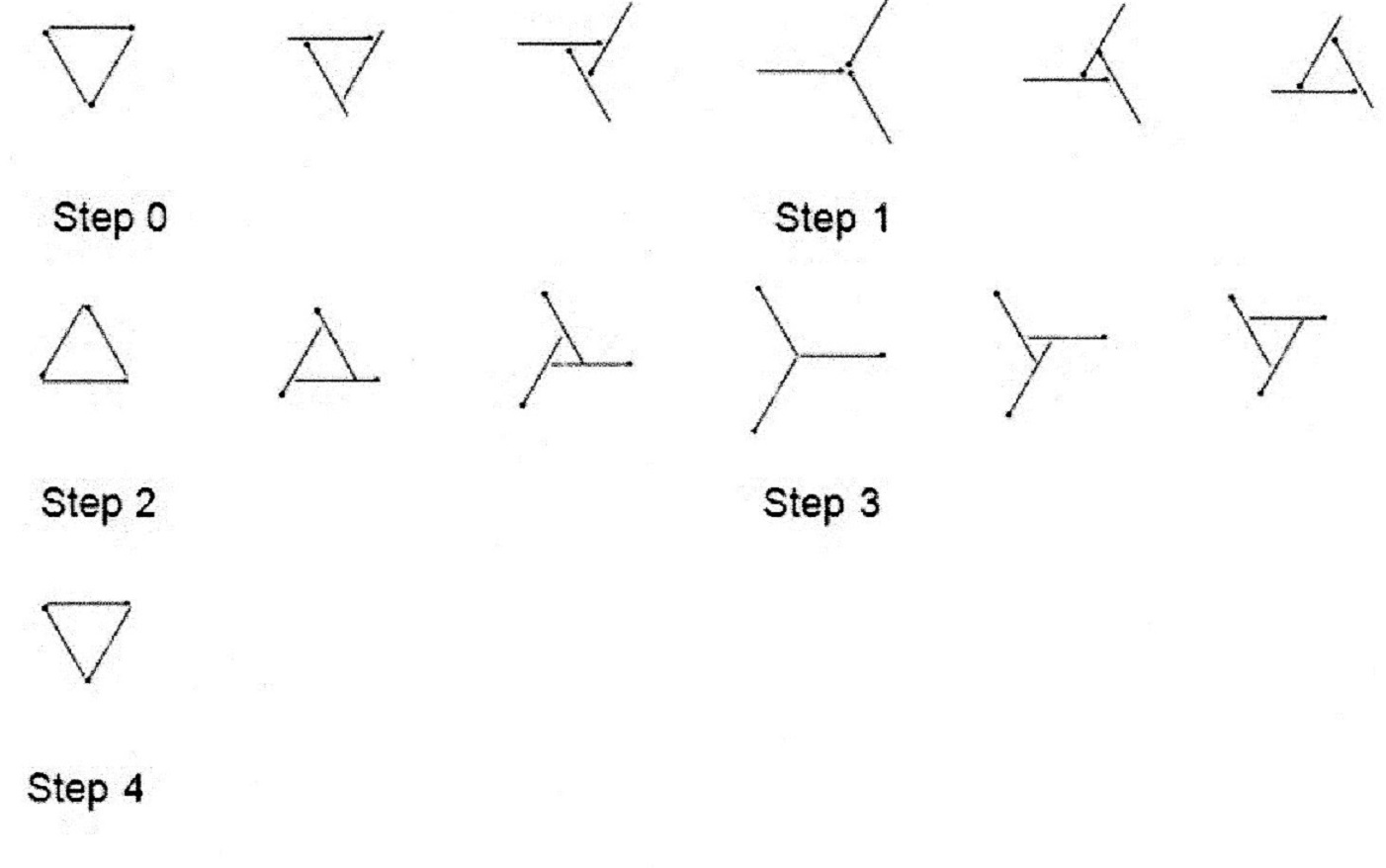} 
\par\end{centering}

\caption{\label{fig:fig5}}
\end{figure}

At the first time-step, the three arcs a, b, and c form a triangle
with the nodes a1b0, b1c0, and c1a0. In Figure \ref{fig:fig5} four
steps are displayed. Intermediate steps show the movements from step
to step.

\qquad{}

\bigskip{}

\subsubsection{Configuration 3: }

\begin{figure}[h]
\noindent \begin{centering}
\includegraphics{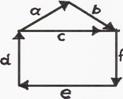} 
\par\end{centering}

\caption{\label{fig:fig6}}
\end{figure}

The initial graph consists of the nodes $\{b_{0},a_{1}\},\{a_{0},c_{0},d_{1}\},\{d_{0},e_{1}\},\{f_{1},e_{0}\},\{b_{1},c_{1},f_{0}\}$
as shown in Figure \ref{fig:fig6},

\noindent \,\ \ \ \ \,\ \ \ \ 

\pagebreak{}The pairs of arcs change their relations according to
the following transition rules:

\noindent \,\ \ \ \ \,\ \ \ \ %
\begin{minipage}[t]{1\columnwidth}%
\noindent $a_{0}c_{0}$\foreignlanguage{ngerman}{$\Rightarrow$} $c_{1}a_{0}$
\begin{description}
\item [{$c_{0}d_{1}$\foreignlanguage{ngerman}{$\Rightarrow$}$d_{0}c_{0}$}]~
\item [{$d_{1}a_{0}$\foreignlanguage{ngerman}{$\Rightarrow$}$a_{1}d_{1}$}]~
\item [{$b_{1}f_{0}$\foreignlanguage{ngerman}{$\Rightarrow$}$f_{1}b_{1}$}]~
\item [{$f_{0}c_{1}$\foreignlanguage{ngerman}{$\Rightarrow$}$c_{0}f_{0}$}]~
\item [{$c_{1}b_{1}$\foreignlanguage{ngerman}{$\Rightarrow$}$b_{0}c_{1}$}]~
\item [{$b_{0}a_{1}$\foreignlanguage{ngerman}{$\Rightarrow$}$a_{0}b_{0}$}]~
\item [{$e_{0}f_{1}$\foreignlanguage{ngerman}{$\Rightarrow$}$f_{0}e_{0}$}]~
\item [{$d_{0}e_{1}$\foreignlanguage{ngerman}{$\Rightarrow$}$e_{0}d_{0}$}]~
\item [{$f_{1}e_{0}$\foreignlanguage{ngerman}{$\Rightarrow$}$e_{0}f_{0}$}]~\end{description}
\end{minipage}

\noindent \,\ \ \ \ \,\ \ \ \ 

\noindent \,\ \ \ \ \,\ \ \ \ 

From the new pairs, the following nodes will result:

$\{b_{0},a_{0},c_{1}\},\{b_{1},f_{1}\},\{a_{1},d_{1}\},\{f_{0}$,
c$_{0}$, d$_{0}$, e$_{0}\}$$ $

\begin{figure}[h]
\noindent \begin{centering}
\includegraphics{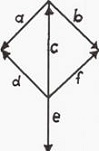} 
\par\end{centering}

\caption{\label{fig:Fig7}}
\end{figure}

The new graph, shown in Figure \ref{fig:Fig7}, is the dual graph
of the initial graph.\\

\begin{figure}[H]
\noindent \begin{centering}
\includegraphics{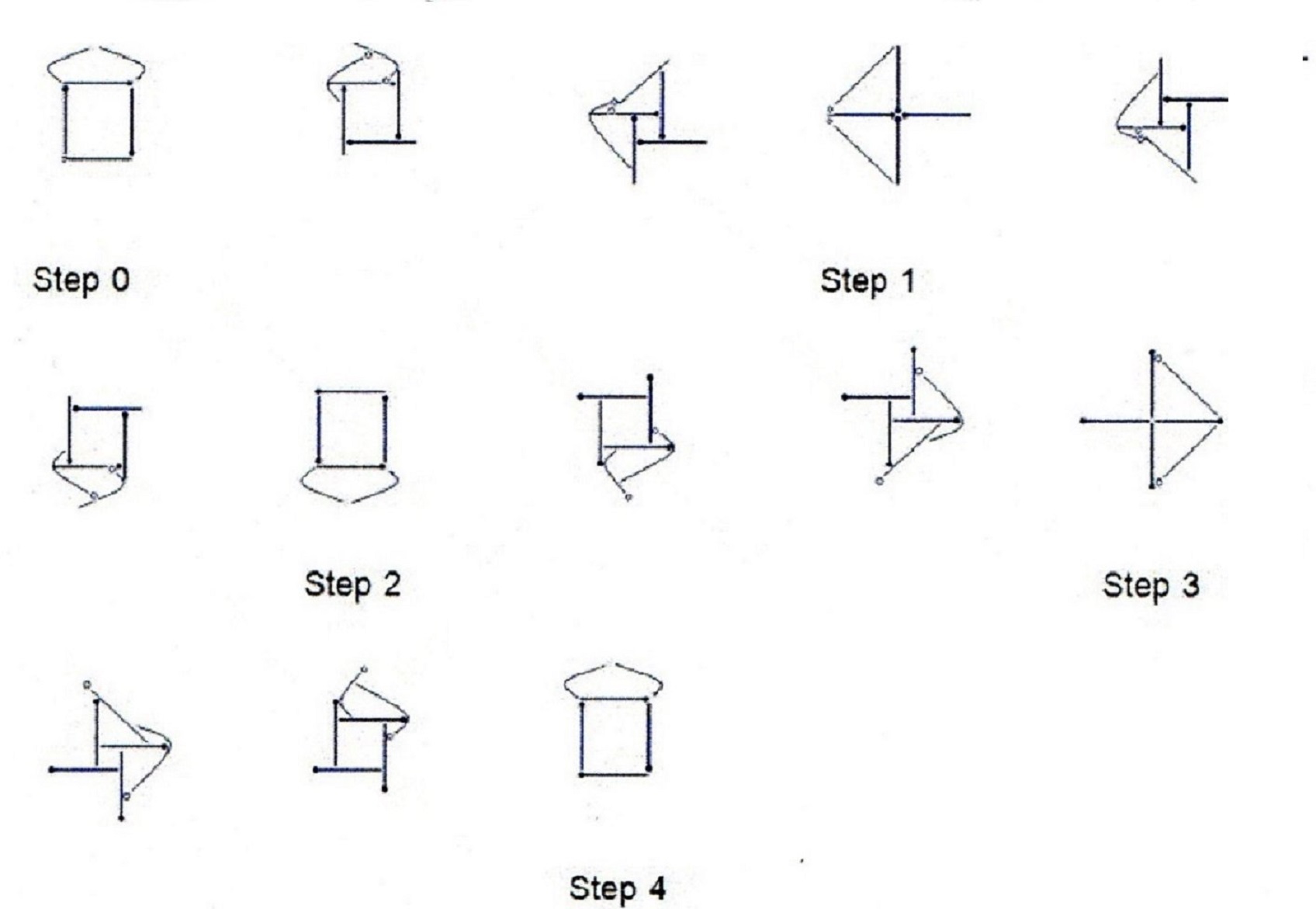} 
\par\end{centering}

\caption{\label{fig:fig8}}
\end{figure}

\noindent \begin{flushleft}
\bigskip{}
 In Figure \ref{fig:fig8}, four steps are displayed. Intermediate
steps show the movements from step to step. 
\par\end{flushleft}

With this rule, planar graphs remain planar after all transitions.
Nonplanar graphs become planar.

\subsection{Rule Example 3 }

The graph consists of colored arcs with the colors $a$, $b$, and
$c$.

The following rules apply for links between arcs of the same color:

$B_{a1}A_{a1}\Rightarrow A_{a0}B_{a1}\Rightarrow B_{a0}A_{a2}\Rightarrow A_{a1}B_{a0}\Rightarrow B_{a1}A_{a1}$

$A_{a1}B_{a1}\Rightarrow B_{a0}A_{a1}\Rightarrow A_{a0}B_{a0}\Rightarrow B_{a0}A_{a1}\Rightarrow A_{a1}B_{a1}$\\

$B_{b1}A_{b1}\Rightarrow A_{b0}B_{b1}\Rightarrow B_{b0}A_{b0}\Rightarrow A_{b1}B_{b0}\Rightarrow B_{b1}A_{b1}$

$A_{b1}B_{b1}\Rightarrow B_{b0}A_{b1}\Rightarrow A_{b0}B_{b0}\Rightarrow B_{b0}Ab1\Rightarrow A_{b1}B_{b1}$

Between arcs of color $c$, there is no interaction. \\

Between arcs of the colors $c$ and $a$, the following transition
rule applies:

$A_{a1}B_{c1}\Rightarrow A_{a1}B_{c0}\Rightarrow A_{a1}B_{c1}$\\

Between arcs of the colors $c$ and $b$, the following transition
rule applies:

$A_{b1}B_{c1}\Rightarrow A_{b0}B_{c1}\Rightarrow A_{b1}B_{c1}$\\

Arcs of the same color other than c can interact with higher priority;
otherwise, the arcs interact with their neighbors by chance.

\paragraph{Configuration: }

According to color, the graph can consist of four arcs, which meet
at a central node.

$a_{a1},b_{a1},c_{a1},d_{a1}$

$e_{b1},f_{b1},g_{b1},h_{b1}$

$i_{c1},j_{c1},k_{c1},l_{c1}$\\
 \\

\begin{figure}[h]
\noindent \begin{centering}
\includegraphics[scale=0.5]{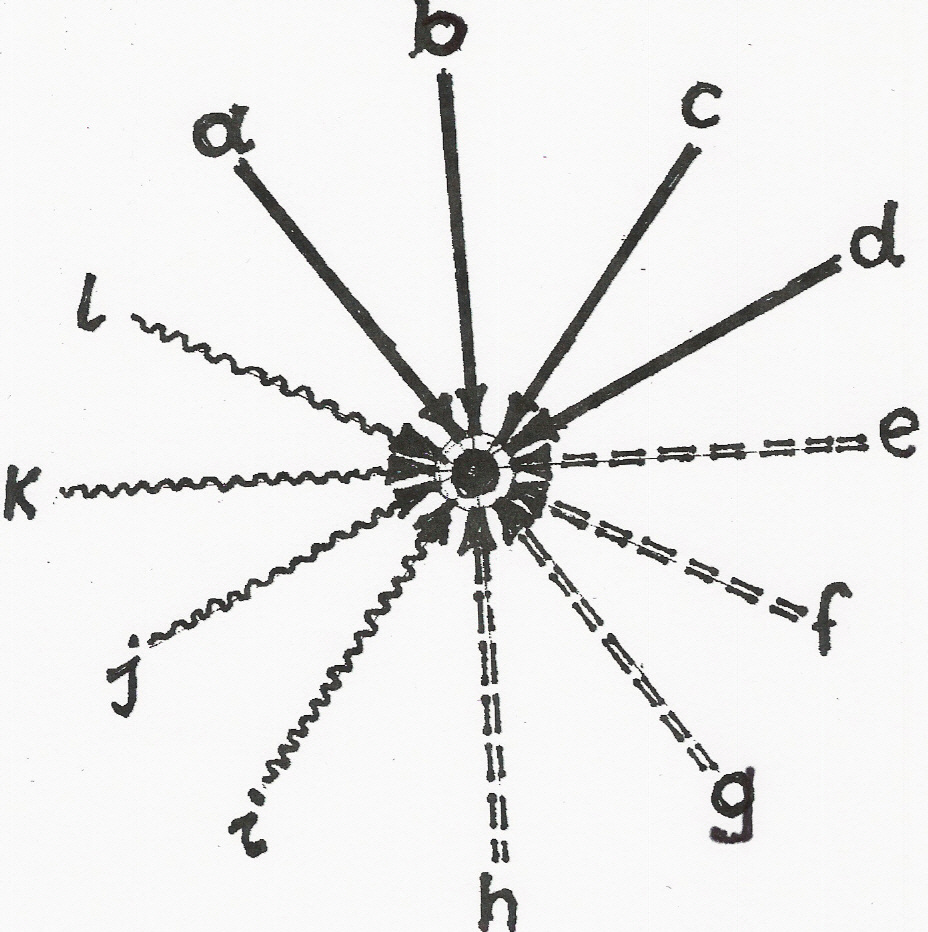} 
\par\end{centering}

\caption{\label{fig:f9}}
\end{figure}

According to the transition rules, for example, the following configurations
can evolve: \\

$a_{a1}b_{a1}\Rightarrow a_{a1}b_{a0}$

$b_{a1}c_{a1}\Rightarrow b_{a1}c_{a0}$

$c_{a1}d_{a1}\Rightarrow c_{a1}d_{a0}$

$d_{a1}a_{a1}\Rightarrow d_{a1}a_{a0}$\\

$a_{a1}i_{c1}\Rightarrow a_{a1}i_{c0}$

$b_{a1}j_{c1}\Rightarrow b_{a1}j_{c0}$

$c_{a1}k_{c1}\Rightarrow c_{a1}k_{c0}$

$d_{a1}l_{c1}\Rightarrow d_{a1}l_{c0}$

\ 

$e_{b1}f_{b1}\Rightarrow e_{b1}f_{b0}$

$f_{b1}g_{b1}\Rightarrow f_{b1}g_{b0}$

$g_{b1}h_{b1}\Rightarrow g_{b1}h_{b0}$

$h_{b1}e_{b1}\Rightarrow h_{b1}e_{b0}$

\ 

$f_{b1}l_{c1}\Rightarrow f_{b0}l_{c1}$

$g_{b1}k_{c1}\Rightarrow g_{b0}k_{c1}$

$h_{b1}j_{c1}\Rightarrow h_{b0}j_{c1}$

$e_{b1}i_{c1}\Rightarrow e_{b0}i_{c1}$\\

\noindent The resulting nodes are as follows: 
\begin{eqnarray*}
\{a_{1},b_{0},i\},\{b_{1},c_{0},j_{0}\},\{c_{1},d_{0},k_{0}\},\{d_{1},a_{0},l_{0}\},\\
\{e_{1},f_{0},l_{1}\},\{f_{1},g_{0},k_{1}\},\{g_{1},h_{0},j_{1}\},\{h_{1},e_{0},i_{1}\}
\end{eqnarray*}

\begin{figure}[H]
\noindent \begin{centering}
\includegraphics{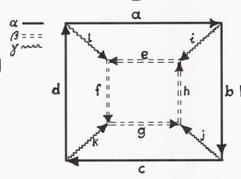} 
\par\end{centering}

\caption{\label{fig:f10}}
\end{figure}

\begin{figure}[H]
\noindent \begin{centering}
\includegraphics{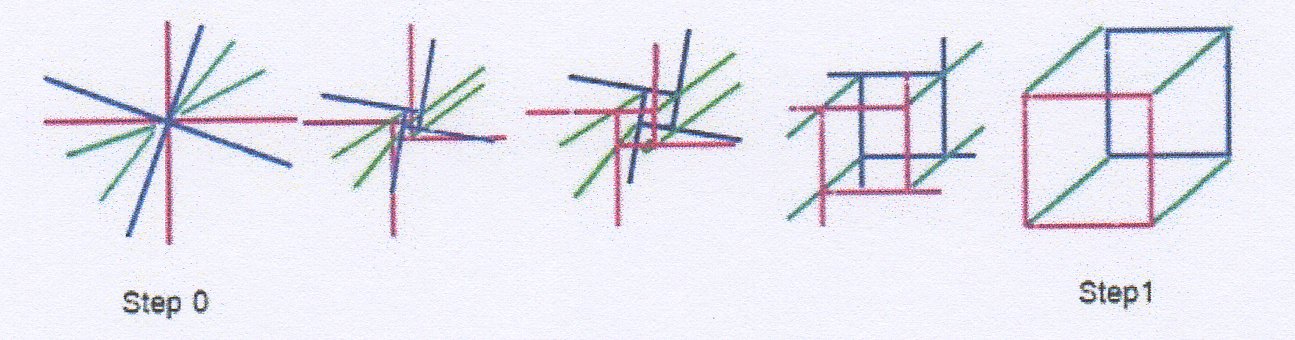} 
\par\end{centering}

\caption{\label{fig:f11}}
\end{figure}

In Figure \ref{fig:f11}, step 0 is topologically equivalent to that
in Figure \ref{fig:f9} and step 1 is equivalent to that in Figure
\ref{fig:f10}. Figure \ref{fig:f11} shows the two steps. Intermediate
steps show the movements from step to step.

\bigskip{}

\begin{figure}[H]
\noindent \begin{centering}
\includegraphics{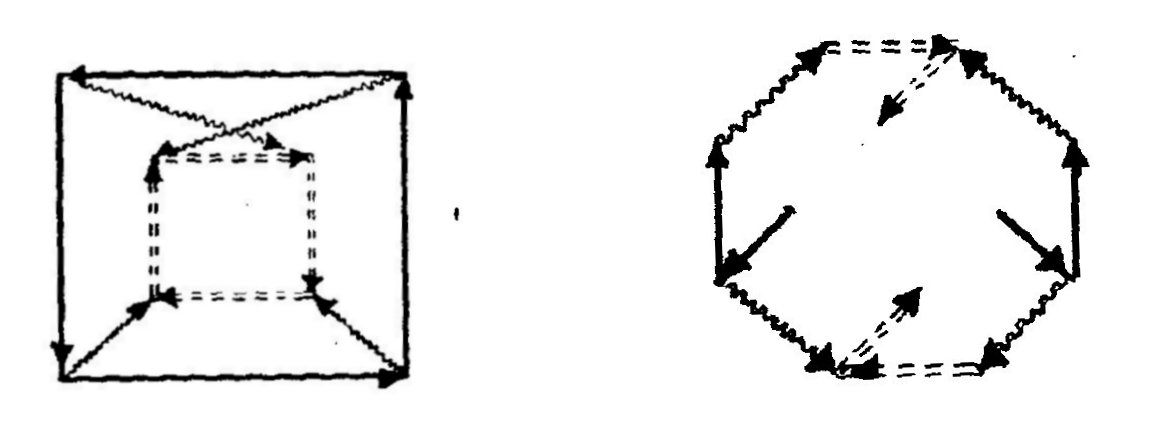} 
\par\end{centering}

\caption{\label{fig:f12}}
\end{figure}

If the arcs interact by chance with their neighbors, independent of
color, results such as those shown in Figure \ref{fig:f12} are possible:

$\:$\bigskip{}

\subsection{Rule Example 4 }

With this example, it can be shown, that a square grid can be created
starting with a set of arcs. The set consists of paired arcs $P$
and normal arcs $N$. A pair $P_{x}$ consists of two arcs, $P_{xa}$
and $P_{xb}$, which are connected at a midpoint $P_{x0}$ (the connection
of $P_{xa0}$ and $P_{xb0}$) and the end points $P_{xa1}$ and $P_{xb1}$.
There is no interaction between the two arcs. If the midpoints of
two pairs meet at the same point, one midpoint $P_{x0}$ moves to
one of the end points of the other pair. Which pair moves to which
end point of the other pair, is arbitrary. A normal arc interacts
with the arcs of a pair as follows: 
\begin{itemize}
\item A normal arc $N_{a}$ interacts alternately with the arcs of a pair
$P_{x}$. When the one end of the arc $N_{a}$ interacts, the other
end is inactive, and vice versa. 
\item If the active end point of $N_{a}$ coincides with the midpoint of
the pair $P_{x}$, then this end point will move to one of the end
points of the pair $P_{x}$. 
\item If the active end point of $N_{a}$ coincides with the end point of
the pair $P_{x}$ and the midpoint of another pair $P_{y}$, then
this active point will move to one of the end points of the new pair
$P_{y}$. 
\item If the active point coincides with an end point of a pair, and this
point is not connected with another pair, then the end point of the
pair will move to the other point of the normal arc $N_{a}$. 
\end{itemize}
The transition rules between two normal arcs might not be relevant
for this example.

\paragraph{Configuration: }

\begin{figure}[H]
\noindent \begin{centering}
\includegraphics{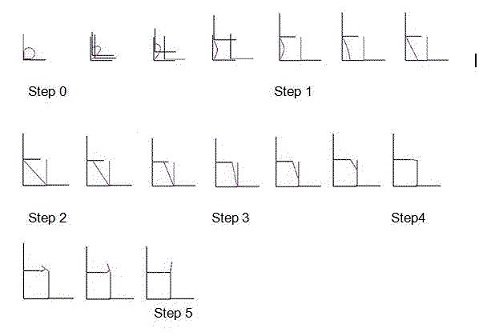} 
\par\end{centering}

\caption{\label{fig:f16}}
\end{figure}

A set of paired arcs and normal arcs meet at a central node. The evolution
to a square grid cell is shown in the series of Fgure \ref{fig:f16},
steps zero through five.

\subsection{Rule Example 5}

With this example, it will be proved that the creation of a cubic
grid starting with a set of arcs is possible. This example is analogues
to Example 4. The set consists of triple arcs $T$ and normal arcs
$N$.

A triple $T_{x}$ consists of three arcs $T_{xa}$, $T_{xb}$, and
$T_{xc}$, which are connected at a midpoint $T_{x0}$, and the end
points $T_{xa1}$, $T_{xb1}$, and $T_{xc1}$. There is no interaction
between the three arcs. If the midpoints of two triplets meet at the
same point, then one midpoint $T_{x0}$ moves to one of the end points
of the other triplet. Which triplet moves to which end point of the
other triplet is arbitrary. The rules regarding how a normal arc interacts
with an arc of a triplet are the same as those in Example 4. \\

\paragraph{Configuration: }

\begin{figure}[H]
\noindent \begin{centering}
\includegraphics{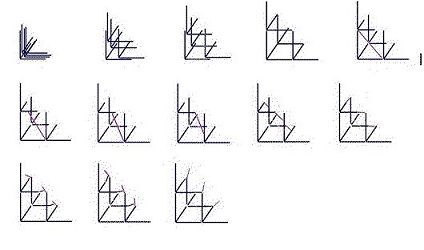} 
\par\end{centering}

\caption{\label{fig:f17-1}}
\end{figure}

The elements of a set of triple and normal arcs meet at a central
node. The evolution to a cubic grid cell is shown in Figure \ref{fig:f17-1}.

\subsection{Rule Example 6 }

It is possible to construct logic gates, such as AND, OR, and NOT,
using an XCA. The information TRUE or FALSE is represented by arcs.
A logic gate consists of three types of arcs: 
\begin{itemize}
\item Information arcs representing information (black) 
\item Conductor arcs for transport of information arcs (blue) 
\item Filter arcs representing filters (red)\\

\end{itemize}
If the start point of an information arc is connected to a conductor
arc, then the information is assumed to be TRUE. If the end point
of an information arc is connected to a conductor arc, then the information
is assumed to be FALSE.

\paragraph{Rules: }
\begin{enumerate}
\item There is no interaction between conductor arcs. 
\item An information arc moves from the end point to the start point of
a conductor arc. 
\item If the start point of a conductor arc encounters an information arc,
but is not connected to any other arc, then the point moves to the
end of the information arc. At the next time step, the information
arc moves back to the end point of the conductor arc. 
\item If two information arcs meet at a conductor arc, a TRUE arc moves
to the start point of a FALSE arc. 
\item If two TRUE arcs meet at a conductor arc, one TRUE arc moves to the
end point of the other arc. 
\item If two FALSE arcs meet at a conductor arc, one FALSE arc moves to
the end point of the other arc. 
\item If one information arc has moved to the other end of the other information
arc, it moves back. 
\item If the arc that moved back meets a filter arc, the information arc
moves to the start point of the filter arc.\\

\end{enumerate}

\subsubsection{Configuration of AND/OR Gate:}

\begin{figure}[H]
\noindent \begin{centering}
\includegraphics[scale=1.5]{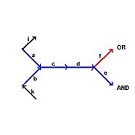} 
\par\end{centering}

\caption{\label{fig:f18}}
\end{figure}

\begin{figure}[H]
\noindent \begin{centering}
\includegraphics{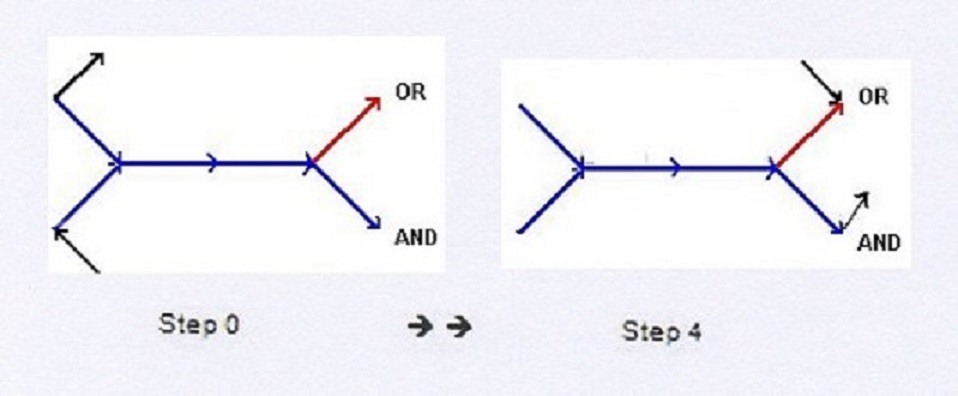} 
\par\end{centering}

\caption{\label{fig:f19}}
\end{figure}

Figure \ref{fig:f18} shows the configuration of the AND/OR gate.
It consists of the conductor arcs $a$, $b$, $c$, $d$, and $e$;
the filter arc $f$; and the information arcs $i$ and $k$. \\
 The time steps evolve as follows: 
\begin{enumerate}
\item The information arcs $i$ and $k$ enter the end points of the conductor
arcs $a$ and $b$. 
\item The arcs $i$ and $k$ move to the start points of a and $b$ (rule
2). Both arcs meet the end point of arc $c$. 
\item The FALSE arc moves to the start point of arc $c$ (rule 2). The TRUE
arc moves to the start point of the FALSE arc (rule 4). If both arcs
are FALSE or TRUE, one arc moves to the other end of the other arc
(rules 5 and 6). 
\item The FALSE arc moves to the start point of arc $d$ meeting the endpoints
of $e$ and $f$. The TRUE arc moves back to the end point of the
FALSE arc (rule 7) and also meets $e$ and $f$. 
\item The TRUE arc moves to the start point of the filter arc $f$ (rule
8). The FALSE arc moves to the start point of $e$ (rule 2). \\

\end{enumerate}
Figure \ref{fig:f19} demonstrates the logic operations with the input
FALSE AND/OR TRUE (at step 0) and the output at step 4 with TRUE for
the OR operation and FALSE for the AND operation.

\begin{figure}[h]
\noindent \begin{centering}
\includegraphics[scale=1.8]{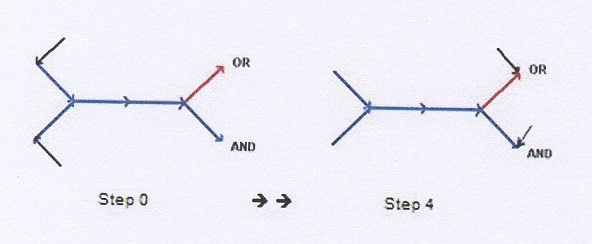} 
\par\end{centering}

\caption{\label{fig:f20}}
\end{figure}

Figure \ref{fig:f20} demonstrates the logic operations with the input
TRUE AND/OR TRUE (at step 0) and the output at step 4 with the result
TRUE for both OR and AND operations. The intermediate steps behave
similarly to the corresponding steps of Figure \ref{fig:f19}.

\subsubsection{Configuration of NOT Gate }

Figure \ref{fig:NOT} shows the configuration of a NOT gate. It consists
of the conductor arcs $a$ and $b$, the filter arc $c$, and the
information arc $i$.\\
 The time steps evolve as follows (see Figure \ref{fig:fig21}): 
\begin{enumerate}
\item The information arc $i$ enters the end point of the conductor arc
$a$. 
\item Arc $i$ moves to the start point of $a$ (rule 2). The arc also meets
the end points of $b$ and $c$. 
\item Arc $i$ moves to the start point of arc $b$ (rule 2). 
\item The start point of arc $b$ moves to the opposite point of arc $i$
(rule 3). 
\item Arc $i$ moves back to the end point of $b$ and encounters the end
point of the filter arc $c$ (rule 8). 
\end{enumerate}
\noindent \begin{center}
\begin{figure}[H]
\noindent \begin{centering}
\includegraphics{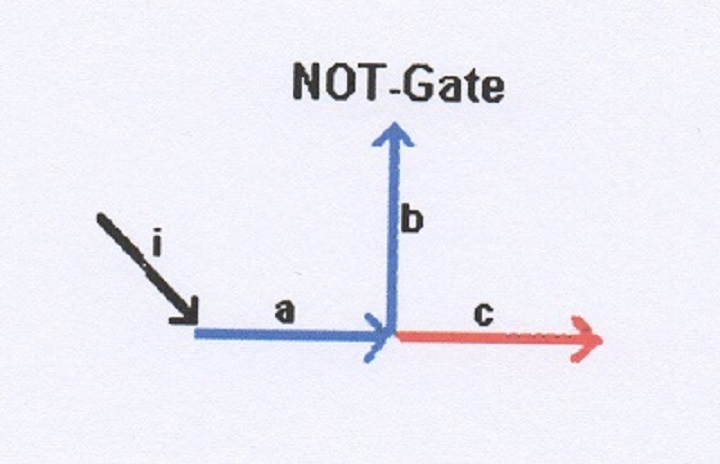} 
\par\end{centering}

\caption{\label{fig:NOT}Not-Gate}
\end{figure}

\par\end{center}

\begin{figure}[H]
\noindent \begin{centering}
\includegraphics{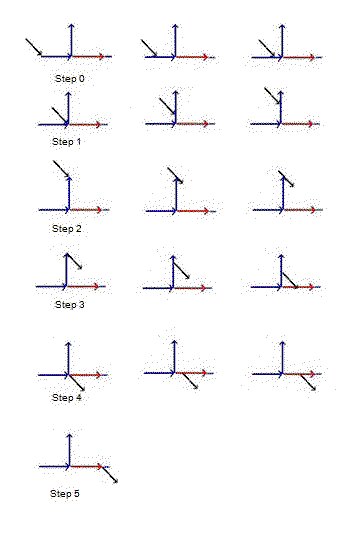} 
\par\end{centering}

\caption{\label{fig:fig21}}
\end{figure}

\subsection{Rule Example 7 - Copy Machine/Oscillator}

The graph consists of three types of arcs: 
\begin{itemize}
\item Several \noun{Type A} arcs to be copied. These arcs have different
properties and do not interact whith each other, except that the arcs
are connected by an arc of \noun{Type C} .
\item A receiver arc of \noun{Type B} .
\item A transmitter arc of \noun{Type C} .
\end{itemize}

\subsubsection{Copy Machine}

\begin{figure}[H]
\noindent \begin{centering}
\includegraphics{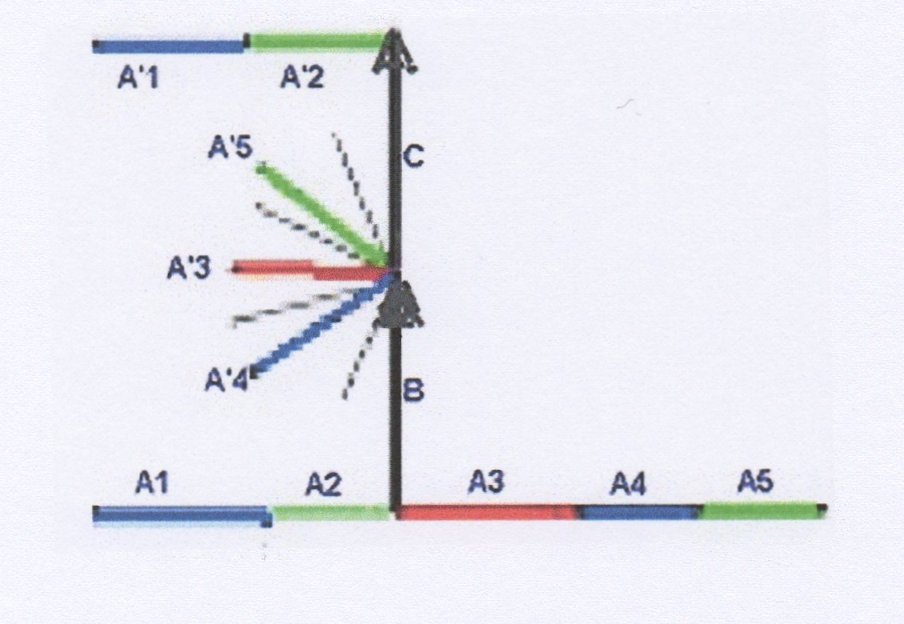} 
\par\end{centering}

\caption{\label{fig:CopyMachine}}
\end{figure}

The foot of the \noun{Type B} arc passes over a path consisting of
\noun{Type A} arcs to be copied (see Figure \ref{fig:CopyMachine}).
The head of the Type B\noun{ }arc is connected to the foot of the
\noun{Type C} arc and a set of \noun{Type A} arcs with different properties.
The head of the \noun{Type C} arc is connected to a path of \noun{Type
A} arcs already copied. If the foot of the \noun{Type B} arc passes
over a \noun{Type A} arc to be copied, it transmits the properties
of this arc to its head and activates an adjacent arc of \noun{Type
A} that has the same properties as that of the arc to be copied. The
activated arc is transmitted from the foot of the \noun{Type C} arc
to its head. The path of the already copied arcs moves to the other
end of the activated arc.

\begin{figure}[H]
\noindent \begin{centering}
\includegraphics[scale=1.5]{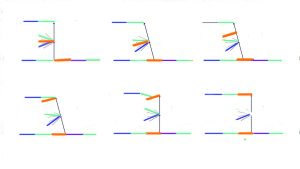} 
\par\end{centering}

\caption{\label{fig:CopyStep1}Step 1}
\end{figure}

\begin{figure}[H]
\noindent \begin{centering}
\includegraphics[scale=1.5]{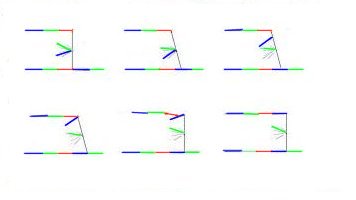} 
\par\end{centering}

\caption{\label{fig:CopyStep2}Step2}
\end{figure}

In Figure \ref{fig:CopyStep1} at step 1, a red arc ($A_{3}$) is
copied ($A\lyxmathsym{\textquoteright}_{3}$); at step 2 (Figure \ref{fig:CopyStep2}),
a blue arc ($A_{4}$) is copied; and at step 3, a green arc ($A_{5}$)
is copied. Intermediate steps show the movements from step to step.

\subsubsection{Oscillator}

The chain of type A arcs can build a loop. The result is then a chain
of copied arcs, which repeate periodically.

\subsection{\label{sub:Rule-Example-9}Rule Example 8 }

The graph consists of four types of arcs: 
\begin{itemize}
\item conductor arcs for transport of Type A, B, or C arcs (black) 
\item \noun{ABC }arcsof\noun{ Type A} (red) 
\item \noun{ABC }arcsof\noun{ Type B} (blue) 
\item \noun{ABC }arcsof\noun{ Type C} (green) 
\end{itemize}

\subsubsection*{Rules: }
\begin{figure}[th]
\includegraphics{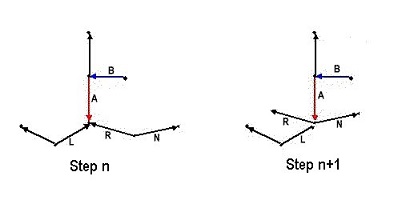}\caption{\label{fig:attract}}
\end{figure}

\begin{figure}[H]
\includegraphics{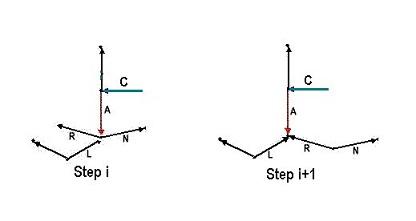}

\caption{\label{fig:repulsion}}
\end{figure}

\begin{enumerate}
\item There is no interaction between arcs of the same type. 
\item IF: The start point $ABC_{1}$ of an arc of type ABC is connected
to a single conductor arc point $L_{i}$ that has no other conductor
arc as neighbor \\
 THEN: With the next step, that start point $ABC_{1}$ will move to
point $L_{k}$, which is the opposite of $L_{i}$ . 
\item IF: The start point $ABC_{1}$ of an arc of type ABC is connected
to two conductor arc points $L_{k}$ and $R_{l}$\\
 and if $ABC_{1}$ was previously connected to the conductor arc point
$L_{i}$\\
 THEN: With the next step, that start point $ABC_{1}$ will move to
point $R_{m}$, which is the opposite of $R_{l}$. 
\item Attraction (see Figure \ref{fig:attract})\\
 IF: The start point $A_{1}$ of the arc of type A was previously
connected to the conductor arc point $L_{i}$ and is now connected
to the opposite point $L_{k}$ and its neighbor $R_{m}$, and the
end point $A_{0}$ of the arc of type A is connected to the start
point $B_{1}$ of an arc of type B. \\
 THEN: With the next step, the start point $A_{1}$ will be connected
to the points $L_{k}$ ,$R_{l}$ , and $N_{m}$, where $R_{l}$ is
the opposite of $R_{m}$, and $N_{m}$ was previously connected to
$R_{l}$ . 
\item Repulsion (see Figure \ref{fig:repulsion})\\
 IF: The start point $A_{1}$ of the arc of type A was previously
connected to the conductor arc point $L_{i}$ and is now connected
to the opposite point $L_{k}$ and its neighbors $R_{l}$ and $N_{m}$,
and the end point $A_{0}$ of the arc of type A is connected to the
start point $C_{1}$ of an arc of type C \\
 THEN: With the next step, the start point $A_{1}$ will be connected
to the points $L_{k}$, and $R_{m}$ where $R_{m}$ is the opposite
of $R_{l}$. 
\item The rules regarding the arc types A, B, and C are also valid if either
A is replaced by B, B is replaced by C, and C is replaced by A, or
A is replaced by C, C is replaced by B, and B is replaced by A 
\end{enumerate}
\newpage{}

\newpage{}

\subsubsection*{Configuration 1}

\begin{figure}[H]
\includegraphics{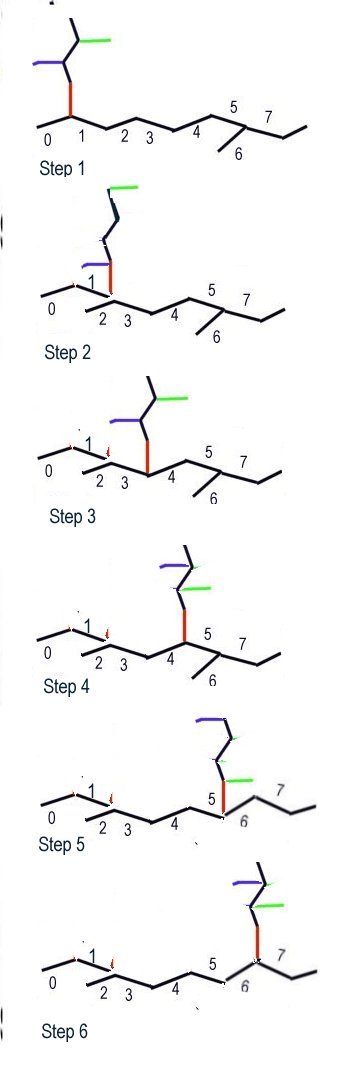}\caption{\label{fig:attract_repulsion}}
\end{figure}

The graph shown in Figure \ref{fig:attract_repulsion} has two chains
of conductor arcs: 
\begin{itemize}
\item A horizontal chain of eight arcs with numbers from 0 to 7 
\item A vertical chain whose lowest arc is linked to the end point $A_{0}$
of an arc of type A\end{itemize}
\begin{description}
\item [{Step}] 1: The start point $A_{1}$ is linked to the arcs 0 and
1 at the left side of the horizontal chain. With the following steps,
$A_{1}$ moves to the next arcs of the horizontal chains from left
to right. The start points $B_{1}$ of an arc of type B and $C{}_{1}$
of an arc of type C are linked to arcs of the vertical chain, as shown
in Figure \ref{fig:attract_repulsion}. 
\item [{Step}] 2: $A_{1}$ moves to the right. $C_{1}$ moves up to the
end of the vertical chain. $B_{1}$ moves down to the bottom of the
vertical chain and is now linked to the end point $A_{0}$ . Owing
to this configuration, rule 4 is applicable. $A_{1}$ is then linked
to the arcs 1, 2, and 3. 
\item [{Step}] 3: $A_{1}$ moves to the right from arc 3 to arc 4. $B_{1}$
moves up and $C_{1}$ moves down according to rule 2. 
\item [{Step}] 4: $A_{1}$ moves to the right from arc 4 to arc 5. $B_{1}$
moves up, and $C_{1}$ moves down according to rule 3. 
\item [{Step}] 5: $A_{1}$ moves to the right from arc 5 to arc 6. $B_{1}$
moves up, and $C_{1}$ moves down according to rule 3. $B_{1}$ moves
up to the end of the vertical chain. $C_{1}$ moves down to the bottom
of the vertical chain and is now linked to the end point $A_{1}$.
Owing to this configuration, rule 5 is applicable. $A_{1}$ is then
linked to arcs 5 and 6. 
\end{description}
\newpage{} 

\subsubsection*{Configuration 2}

The vertical chain of conductor arcs shown in Figure \ref{fig:attract_repulsion}
consists of only three arcs. This chain can be extended by any number
of conductor arcs.

\subsubsection*{Configuration 3}

The configurations can be extended using additional chains in combination
with ABC arcs. The end point $B_{0}$ (blue arc), for instance, can
also be linked to a chain of conductor arcs. The start points of two
arcs of types A and C are then connected to this chain. In this way,
a nested system of conductor arcs and ABC arcs can be constructed.

\subsubsection*{Configuration 4}

\begin{figure}[th]
\includegraphics{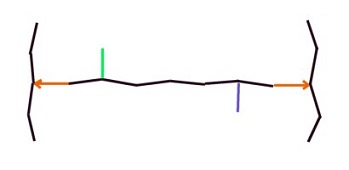}

\caption{\label{fig:photon}}
\end{figure}

In Figure \ref{fig:attract_repulsion}, the upper end of the vertical
chain is not connected to any arc. In Figure \ref{fig:photon}, the
horizontal chain with type A, B, and C arcs is terminated by type
A arcs at both ends. Therefore, this chain will move on both sides
along the vertical chains of conductor arcs at the same time. The
type B and C arcs travel back and forth between the ends of the horizontal
chain. The frequency of this commuting depends on the chain length.

\subsubsection*{\newpage{}}

\section{Signals, Information Arcs, ad Information Pulses }

In Rule Example 6, the information TRUE or FALSE is represented as
an information arc. As an alternative, the values $V(A_{0})/V(A_{1})$
can be used for transmitting an information pulse {[}the values $V(A_{0})$
and $V(A_{1})$ are assigned to the start and end points, respectively{]}.

Let us consider the sequence of arcs $a$, $b$, and $c$, where $a_{1}$
meets $b_{0}$ and $b_{1}$ meets $c_{0}$. A signal (information
pulse) starts at $a_{0}$ and propagates to $a_{1}=b_{0}$, then to
$b_{1}=c_{0}$, and then to $c_{1}$ from step to step. According
to cellular rules, the signal can change from step to step. With information
pulses instead of information arcs, the configurations of the AND/OR
gates (Example 6), and the Copy Machine (Example 7) can be constructed
more simply.

The AND gate performs a logical AND operation on two logic inputs
and produces a single logic output. The end points of the two conductor
arcs $a$ and $b$ receive information pulses at $a_{0}$ and $b_{0}$.
These pulses are propagated to their start points $a_{1}$ and $b_{1}$.
The start points meet at the end point $c_{0}$ of the filter arc
$c$. If both information values at the node \{$a_{1},b_{1},c_{0}$\}
are TRUE, then a signal for TRUE is transmitted from the end point
$c_{0}$ to the start point $c_{1}$. If one of the information values
is FALSE, then a signal for FALSE is transmitted. 

The Copy Machine copies the properties of the nodes of a chain with
the arcs $a,b,c,d...$ to the nodes of a parallel chain with the arcs
$a'$, $b',c',d'...$ In the start, a copier arc is linked to the
endpoints $a_{0}$ and $a'_{0}$, and transmits the value $V(a_{0})$
to $V(a'_{0})$. Then, the copier arc moves to the start points of
$a$ and $a'$ and copies $V(a_{1})$ to $V(a'_{1})$. With the next
step, the copier arc moves to $b_{1}$ and $a'_{1}$, and so forth.
The abovementioned examples describe how information pulses can travel
through a graph without changing the graph structure. Even more interesting,
but also more complex, are systems in which the interaction of arcs
can produce information pulses, which then trigger the interaction
of arcs again.

\newpage{}

\section{\label{sec:Visualization}Visualization }

The pularity of CAs might be a consequence of the following facts:

\textbullet{} \noun{CA} is often used to simulate natural or mathematical
phenomena that can be mapped on a plane or space.

\textbullet{} The grid can be mapped onto any medium, such as paper,
computer memory, or monitors.

\textbullet{} The grid has a uniform pattern, and the neighbors of
a cell do not change their positions.

\textbullet{} The rules can be executed easily for each cell.

Visualizing \noun{XCA} games is not as easy as visiualizing CA games,
because the neighbors repeatedly change and their graphs are difficult
to visualize. All graphs shown above are drawn manually in such a
way that the graphs look rather intelligible. However, drawing very
large graphs manually is nearly impossible. In general, there is no
unique procedure for drawing a graph, nor is there a unique algorithm
for visualizing a graph. For example, you can draw a graph by assigning
to each vertex any arbitrary coordinate within a given frame. The
result can then look rather chaotic. Drawing the graph more meaningfully
requires adequate algorithms. An Internet search for \textquotedblleft graph
drawing\textquotedblright{} returns nearly one million matches, including
many algorithms for graph visualization. For drawing the graph of
the first time step of an \noun{XCA} game, force-based algorithms
(also known as force directed placement, or force directed layout
algorithms) appeare to be the best choice \cite{key-15}.

\footnote{\textquotedblleft The force-directed algorithms achieve this by assigning
forces amongst the set of edges and the set of nodes. The most straightforward
method is to assign forces as if the edges were springs (Hooke's law)
and the nodes were electrically charged particles (Coulomb's law).
The entire graph is then simulated as if it were a physical system.
The forces are applied to the nodes, pulling them closer together
or pushing them further apart. This is repeated iteratively until
the system comes to an equilibrium state; i.e., their relative positions
do not change anymore from one iteration to the next. At that moment,
the graph is drawn. The physical interpretation of this equilibrium
state is that all the forces are in mechanical equilibrium.\textquotedblright{} }

Animations of the examples, shown above, are designed manually in
such a way as to show how the graphs of consecutive time steps evolve.
This is achieved by inserting intermediate steps, as was explained
in Rule Example 2 using Figure \ref{fig:fig4}.

\section{Artificial World}

Using an \noun{XCA}, an artificial world can be simulated, beginning
with a \emph{Big Bang}.

According to the Big Bang theory, our universe began with a singularity.
All energy was concentrated in one singular point. Space did not exist
from the beginning. Since physical theories presupposes the existence
of space, time, and matter, physical theories were not valid before
the creation of space. How was space created after the Big Bang and
what happened between the starting time and the time when space came
into existence? Nobody knows, and imagining what happened is difficult.

There are many approaches to solving this mystery. The existence of
Plank's length and time indicates that space has a granular structure,
rather than being a continuum. Some authors, such as Zirek\cite{key-19}
and Ostoma and Trushyk\cite{key-18}, have attempted to explain the
Big Bang and the universe as a CA consisting of a huge array of cells
capable of storing numeric information. They might be able to explain
physical phenomena using CA. However, this approach presupposes the
existence of a grid from the very beginning, and a grid is not a singularity.

Stephen Wolfram believes that space is a giant network of nodes\cite{key-16}\cite{key-17}.
It could work similarly to a so-called substitution system by which
some node or substructure of the network could be replaced at each
time step by another structure according to some fixed rule. The substitution
system requires the neighborhood of the replaced structure, and the
substitution remains unchanged.

With this assumption, the creation of a grid can be explained, starting
from a singular node. However, only a planar grid can evolve in this
system, and the network will grow endlessly.

I follow up on the idea that space is a giant network. However, I
am considering the XCA model with which I am trying to explain the
start of the universe and the creation of space by the evolution of
a dynamic graph.

Many people, including me, believe that the universe is a giant computer
system. The subject of this belief is known as \emph{digital physics}.

Many different worlds can be created with the many possible XCA rules.

I am tempted to believe, that with XCA, it will be possible to simulate
the physical world, provided that appropriate rules are found. This
is not yet the case.

\subsection{Space/Grid }

Stephen Wolfram believes \textquotedblleft that what is by far the
most likely is that the lowest level space is a giant network of nodes\textquotedblright \cite{key-16}\cite{key-17}.
I also believe that a giant XCA graph (network) with appropriate rules
and configuration can create its own space.

Graphs have properties similar to those of space, but unlike space,
graphs are never continua. However, grid graphs (lattice graphs, tilings,
tessellations) can be considered as a substitude for space. In contrast
to CAs, an XCA does not require a grid, but it can create one, just
as the real universe after the Big Bang generated its own space, which
had not existed previously. Rule Examples 4 and 5 above demonstrate
that the creation of grid graphs is possible. The grid can be a square
grid or a cubic grid, but it can also be an irregular (or unstructured)
grid. A planar graph can be considered as a two-dimensional irregular
grid graph. In general, d-dimensional irregular grid graphs consist
of d-dimensional cells bounded by (d -1)-dimensional faces. The cells
and faces must not be intersected by any object.

\footnote{My conjecture is that d-dimensional irregular grid graphs exist if
they have a dual graph. In this sense a cell is assigned to a vertex
and a vertex is assigned to a cell. With a cubic grid, for instance,
a cell (solid object) can be replaced by a vertex and a vertex by
a cell, and the faces and edges are to be interchanged as well.}

\subsection{Attraction/Repulsion}

In the real world, the distance between objects can decrease or increase
as a result of forces.

Let us consider an arc $a$ that is connected to the start points
of the arcs $b$ and $c$. The distance between the end points of
the arcs $b$ and $c$ is either two, if all arcs are neighbored,
or three, if $b$ and $c$ are not neighbored. Transition from the
constellation with a distance of two to the other constellation can
be seen as repulsion, whereas the transition from distance three to
distance two can be seen as attraction.

Figure \ref{fig:attract} of Rule Example \ref{sub:Rule-Example-9}
demonstrates how attraction might work.

Figure \ref{fig:repulsion} demonstrates how repulsion might work.
The chain of conductor arcs with L and N is enlarged by the arc R.
If the third arc R does not exist, then there is no enlargement. Normally,
the balance between attraction and repulsion might be retained. However,
if in some cases, the enlargement does not occur, then attraction
will outweigh repulsion.

In reality, attraction outweighs repulsion through the gravitational
force. The counter force of gravitation is the centrifugal force.

In the XCA world, the centrifugal force might be caused by a restriction
rule to the effect that transitions must not allow reversing, which
means that the status of two arcs at time step t +1 is the same as
that time step t -1.

\subsection{Virtual Dimension}

The considerations above apply to grids of a discrete time step. If
we observe the evolution, over two or more steps, we might find that
the dimension of a grid is higher than that of the grid per step.
Example 1: We know that a complete graph with five nodes is not planar.
If we discard one of the 10 arcs, then the graph is planar. Let us
assume that with a next time step, another arc is removed from the
complete graph. The graphs of both steps are planar. However, if we
overlay the steps, the graphs look as if they are complete.

\begin{figure}[th]
\includegraphics{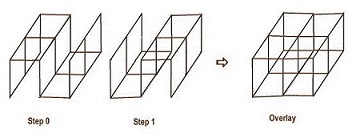}

\caption{\label{fig:VirtualDimension}Virtual Dimension}
\end{figure}

Example 2: The graphs of steps 0 and 1 In Figure \ref{fig:VirtualDimension}
are planar, while the overlay of both graphs is a cubic grid.

\subsection{Time Aspects}

\subsubsection{Step-based Time }

With an XCA, time can be measured by the number of steps, provided
that the time between two steps is the same for all arcs, i.e., all
arcs interact in a synchronized manner. However, what might happen
if the arcs are not synchronized? Let us consider two neighboring
arcs with different time steps. The arcs might then interact, for
instance, according tothe following rule examples: 
\begin{itemize}
\item The faster arc must wait until the slower one is ready. 
\item The active arc predicts the time of interaction. 
\item The arcs interact only when a signal arrives. 
\end{itemize}

\subsubsection{Event Time }

If we compare two areas of a graph at a particular time span (number
of steps), then we might find that one area has fewer transactions
than the other. This could be interpreted such that the time in the
first area is slower than that in the other one. In an extreme case,
where no changes occur in the first area, time stands still.

\subsubsection{Space Time }

The length between two nodes can be measured by the number of arcs
that build the path. However, this number changes from step to step.
A better measurement of the distance between two nodes might be the
shortest time that an arc or signal requires to travel from one node
to the other one. Note, that with this definition, the distance measured
from node $A$ to node $B$ is not necessarily identical with that
measured from node $B$ to node $A$. Let us assume two nodes $A$
and $B$. Many different paths connect the nodes. At the start, many
signals are spread from node $A$ over the system. The signals move
randomly through any path, and some might arrive at node $B$ earlier
or later. The first signal that arrives at B has then traversed the
shortest path between $A$ and $B$ and determines the distance between
$A$ and $B$. The shortest path of the next time step is not necessarily
the same as that of the previous one.

\subsection{Elementary Particles}

The crucial statement of this hypothesis is that all phenomena, succh
as space, radiation, and particles, originate from the different configurations
and evolutions of subgraphs. 
\begin{itemize}
\item Empty space, that is, an area free of radiation and particles, nevertheless,
contains arcs. Because arcs are considered to have energy units, empty
space aldo contains energy. 
\item A photon, emitted by a particle, moves with the highest speed to another
particle, by which it is absorbed. The chains with type A, B and C
arcs, as described in Chapter \ref{sub:Rule-Example-9}, might serve
as a model of a hypothetical photon. The length of the chain defines
the frequency. Both ends of the chain can move to different destinations. 
\item Elementary particles are more complex and have long lives. Interactions
with the environment cannot destroy their identity. A hypothetical
particle might consist of a kernel surrounded by a planar surface.
Arcs of the surface are connected with arcs of the environment. However,
there are no direct connections between the environment and the kernel.
While the environment is a three-dimensional grid, the kernel is more
complex and can consist of higher dimensions. 
\end{itemize}

\section{Conclusion }

The examples with different rules show how manifold the concept of
XCA is. However, in addition to the promising results, there are still
open problems. Just as the Game of Life simulates birth, survival,
and death, an XCA game can simulate a system that starts from a singularity,
and evolves into a complex space. Although an example proves that
this is possible, that example is not comparable with the versatility
of the Life Game. No example of this article is shown with more than
16 arcs and 6 steps. Obtaining more meaningful results, requires more
cells (arcs) over many time steps. This is true for both CAs and XCAs.
As stated in Chapter \ref{sec:Visualization}, drawing very large
graphs manually is nearly impossible. To explore the behavior of very
large XCA systems over a huge number of steps, we need adequate programs,
that calculate and visualize the consecutive steps. I would not have
published this article before having such programs. However, I am
more than 80 years ({*}1933) old, and I do not have much time left
to come to a reasonable end. Now, it might be your turn to continue
the game.

\section{Acknowledgement}

The author would like to thank Journal Prep (\href{http://www.journalprep.com}{www.journalprep.com})
for a peer review, and to thank Enago (\href{http://www.enago.com}{www.enago.com})
for the English language review.

\newpage{}

\appendix

\section{Appendix}

\subsection{\label{sub:Cellular-Automaton}Cellular Automaton}

In 1970 John H. Conway published ``The Game of Life\textquotedblright{}
\cite{key-7} that became soon very popular. This ``game\textquotedblright{}
is the best-known example of a cellular automaton \cite{key-8} (CA).
`The Game of Life (or simply Life) is not a game in the conventional
sense. There are no players, and no winning or losing. Once the \textquotedbl{}pieces\textquotedbl{}
are placed in the starting position, the rules determine everything
that happens later' \cite{key-9}. `The \textquotedbl{}game\textquotedbl{}
is a zero-player game, meaning that its evolution is determined by
its initial state, requiring no further input from humans. One interacts
with the Game of Life by creating an initial configuration and observing
how it evolves.' \cite{key-10}

Life is an example of self-organization \cite{key-11}.

The different examples of the cellular automaton show how chaos may
evolve into a structured organization.

See some examples on the site: \href{http://www.collidoscope.com/modernca/}{http://www.collidoscope.com/modernca/}
\cite{key-12}. Before going into details, I am introducing the general
concept of cellular automaton.

A CA consists of an n-dimensional grid, to the cells of which certain
values are assigned at a definite time. For each next time step, the
value of each cell changes according to the values of the neighboring
cells (usually including the cell itself) and according to some transition
rules.

\begin{figure}[th]
\includegraphics{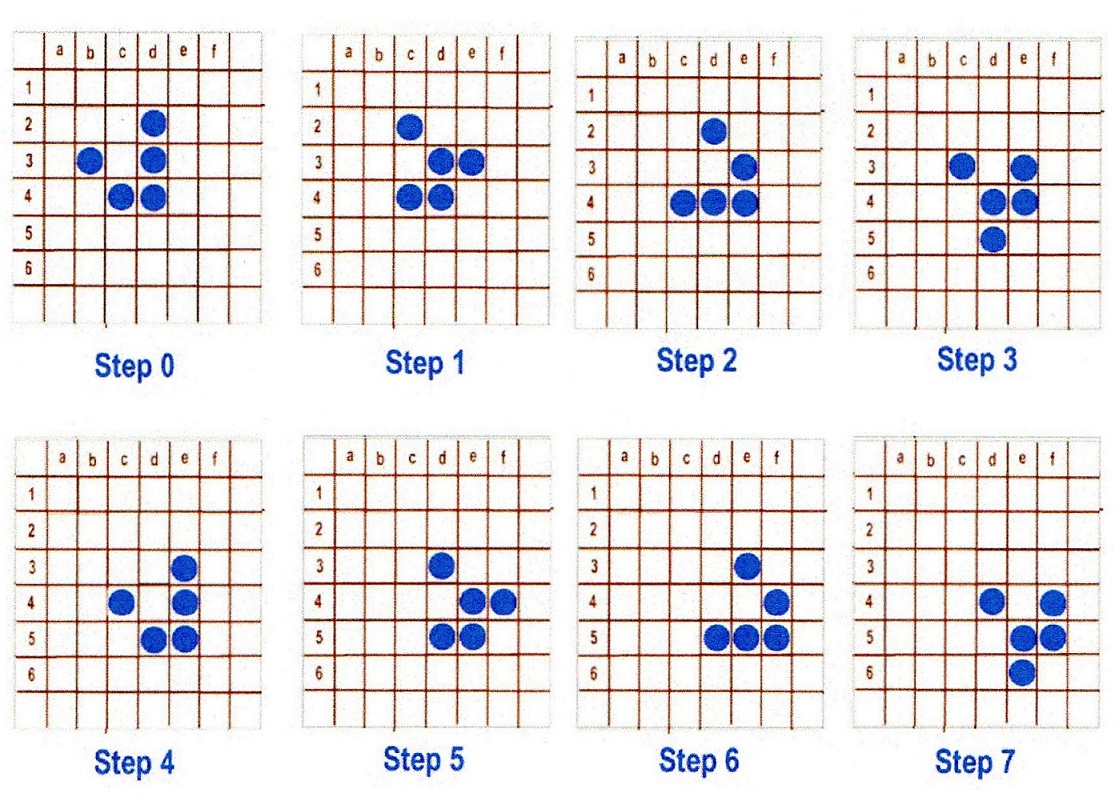}

\caption{\label{fig:Glider}Glider}
\end{figure}

As an example, the Game of Life acts on a two-dimensional rectangular
grid. Each cell can be in one of two states; live or dead (represented
by the values zero and one). A square cell has eight adjacent neighbors,
four orthogonal and four diagonal cells. At each time step the following
transitions occur simultaneously to every cell: 
\begin{itemize}
\item A dead cell with exactly three live neighbors becomes a live cell
(birth) 
\item A live cell with two or three live neighbors stays alive (survival) 
\item In all other cases, a cell dies or remains dead (overcrowding or loneliness)\\

\end{itemize}
Figure \ref{fig:Glider} shows a so-called Glider. After every fourth
step, the patterns are repeating. However, this patterns are shifted
by one row and column, thus gliding diagonal from top left to bottom
right. Let us see how the starting configuration of step 0 evolves
to step 1. 
\begin{enumerate}
\item All cells in 1st row: these cells have no three live cells; therefor
remain dead 
\item All cells in 1st column: these cells have no three live neighbors;
therefor remain dead 
\item Live cell on 2nd column (b), 3rd row (3): this cell has neither two
nor three live neighbors; therefor gets dead 
\item All dead cells in 2nd column remain dead because they have no three
live neighbors 
\item Dead cell on 3rd column (c), 2nd row (2): this cell has three neighbors;
therefor gets alive 
\item Dead cell on 3rd column (c), 3rd row (3): this cell has five neighbors;
therefor remains dead 
\item Live cell on 3rd column (c), 4th row (4): this cell has three live
neighbors; therefor remains alive 
\item Live cell on 4th column (d), 3rd row (3): this cell has only one neighbor;
therefor gets dead 
\item Live cell on 4th column (d), 4th row (4): this cell has two live neighbors;
therefor remains alive 
\item Dead cell on 5th column (e), 3rd row (3): this cell has three live
neighbors; therefor gets alive 
\item All remaining dead cells remain dead because they have no three live
neighbors 
\end{enumerate}

You may find many sites about CA and the Game of Life on the World
Wide Web \cite{key-14}.

Even if the game starts with chaotic condtitions, after a number of
time steps different types of patterns may be seen: 
\begin{itemize}
\item `still lives' are patterns that do not change 
\item `oscillators' repeat their status 
\item Sometimes patterns appear that seem to replicate themselves. The `glider'
shown in Figure\ref{fig:fig2} is such a pattern 
\end{itemize}

\subsection{Adjacency Matrix, Incidence$ $ Matrix, Incidence$ $ List}

A graph can be described by an Adjacency Matrix. Normally, with the
classical view, this matrix shows how the nodes are connected; when
the entry at row, column is 1 in the matrix, the nodes are connected.
With the XCA view, where arcs are seen as objects, the adjacency matrix
represents which end points of arcs are adjacent to which other endpoints
of arcs. The adjacency matrix on $n$ arcs is then a $2n\times2n$
matrix, (or a $n\times n$ matrix, the elements of which are $2\times2$
sub matrices).$ $ If the end point $a_{i}$ of arc $a$ meets the
end point $b_{j}$ of arc $b$, then the entry of $a_{i}b_{j}$ is
1 else the entry is 0.

\begin{figure}[H]
\noindent \begin{centering}
\includegraphics[scale=0.4]{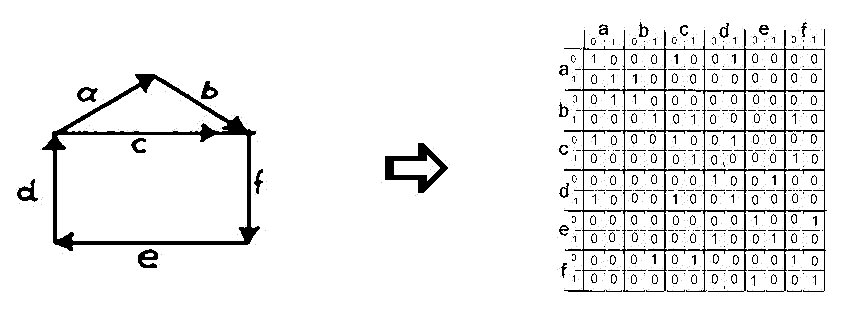} 
\par\end{centering}

\caption{\label{fig:Adjacency-Matrix}Adjacency Matrix}
\end{figure}

In Figure \ref{fig:Adjacency-Matrix}, for instance, the entry of
$e_{1}d{}_{0}$ is 1.

Another matrix representation for a graph is the incidence matrix.
This matrix is the same for both views.

The incidence matrix assigns each row to a node and each column to
an edge. For a standard incidence matrix a 1 appears wherever a row's
node is incident on the column's arc. The elements are zero otherwise.

The incidence matrix can be compressed to the incidents' list: to
each arc, the source and target node is assigned.

\subsection{PSEUDO CODES OF THE BIG BANG GAME}

The following pseudo codes do not reflect the performance problems.
Since the matrix elements are ones and zeros, these values can be
packed within bytes and may be considered as 2 x 2 sub matrices. The
bits in a sub matrix can then be addressed as binary numbers.

For instance, the bit in the 2 x 2 sub matrix S{[}0{]}{[}1{]} can
be represented by the decimal number 0 x 1 + 1 x 2 + 0 x 4 + 0 x 8
= 2,

or S{[}1{]}{[}0{]} by 0 x 1 + 0 x 2 + 1 x 4 + 0 x 8 = 4

\subsubsection{BIG BANG GAME}

\begin{lstlisting}
//
//
//Input: Incidents List L, number of arcs n, number of nodes m, 
//		Property List P with properties assigned to arcs
//		Rules R
//		number of time steps nt
begin
  read L,n,m,P,nt
  for each step from 0 to nt
    //Convert incidence list to adjacency matrix
      call Incidence2Adjacency(L,A,n,m)
    //Apply transition rules on adjacency matrix
      call Transition(A,B,n,P,R)
    //Convert adjacency matrix to incidence matrix
      call Adjacency2Incidence(A,n)
    //Convert incidents matrix to incidence list
	  call IncidentsMatrix2List(A,L,n,m)
      Visualise the graph
  end for
end
\end{lstlisting}

\subsubsection{Transition}

\begin{lstlisting}
procedure Transition(A,B,n,P,R)
//Input: Adjacency Matrix A with n Arcs //first row and column are starting with index = 1
//Output: Matrix B
begin							//1
row = 1
column = 2
for each row from 1 to n	//2
  for each column from 2 to n	//3
	calculate new value B[row][column] 
	according to the transition rules dependent on A[row][column] 
  end for				 //3 - column
end for					//2 - row
return matrix B
end						 //1
\end{lstlisting}

\subsubsection{CONVERT INCIDENCE LIST TO ADJACENCY MATRIX }

\begin{lstlisting}
procedure Incidence2Adjacency(L,A,n,m)
//Input:Incidents List with n arcs and m nodes 
//first arc and node are starting with index = 1
//Output:Adjacency Matrix A. 

Incidents List L(arc,source,target)
Incidents Matrix A
Node List NodeList[n][m]
SourceList[m][n]
TargetList[m][n]
begin								//1
node = 0
for each arc in L from 1 to n		//2
  GetIncidentsList(arc,source,target)
  SourceList[source][arc] = 1
  TargetList[target][arc] = 1
end for							   //2
Create new matrix A[2][2][n][n]
for each node from 1 to m			 //3
  arcsource = 0
  arctarget = 0
  for each arc from 1 to n			//4
    if SourceList[node][arc] is TRUE  //5
    then
      if arcsource is 0
      then 
        arcsource = arc
      end if
    end if							//5
    if TargetList[node][arc] is TRUE  //6
    then
      if arctarget is 0
      then 
        arctarget = arc
      end if
    end if							//6
    if arcsource < arctarget		  //7
    then 
      arc_node = arcsource
      sub_node = 0
    else
      arc_node = arctarget
      sub_node = 1
    end if
  end for							 //7
  //
  if sub_node = 0
  then
     sub_alt = 1
  else
     sub_alt = 0
  end for
  for each arc from 1 to n			//8
    if SourceList[node][arc] is TRUE
    then
      SetAdj(A,n,sub_node,0,arc_node,arc,TRUE)
      SetAdj(A,n,sub_alt, 1,arc,arc_node,TRUE)
    end if
    if TargetList[node][arc] is TRUE
    then
      SetAdj(A,n,sub_node,1,arc_node,arc,TRUE)
      SetAdj(A,n,sub_alt ,0,arc,arc_node,TRUE)
    end ifCONVERT INCIDENCE LIST TO ADJACENCY MATRIX 
  end for							 //8
end for							   //3
//Completion of the adjacency matrix
//If arc i is adjacent to j and arc arc j is adjacent to k
//then arc i is also adjacent to k
for each row from 1 to n	//9
  for each column from 1 to n	//10
    for each subrow from 0 to 1  //11
      for each subcolumn from 0 to 1	//12
        //
        if GetAdj(A,n,subrow,subcolumn,row,column) is TRUE	//13
        then
          for each row1 from 1 to n		//14
           for each subrow1 from 0 to 1		//15
            if GetAdj(A,n,subrow1,subcolumn,row1,column) is TRUE	//16
            then
             for each column1 from 1 to n		//17
              if GetAdj(A,n,subrow1,subcolumn,row1,column) is TRUE	//18
              then
               SetAdj(A,n,subrow, subcolumn,row, column1,TRUE)
              end if //18
             end for //17 - column1
            end if   //16
           end for   //15 - subrow1
          end for    //14 - row1
        end if       //13 - GetAdj
        //
      end for  //12
    end for                      //11
  end for						//10
end for	//9
end 	   //1
\end{lstlisting}

\subsubsection{CONVERT ADJACENCY MATRIX TO INCIDENCE MATRIX}

\begin{lstlisting}
procedure Adjacency2Incidence(A,n)
//Input: Ajacency Matrix A with n Arcs //first row and column are starting with index = 1
//Matrix A, converted to Incidents Matrix. Some of the 2 x n nodes are isolated
//Output: Incidence List: source and target nodes assigned to arc
//Enumerate subrows from 1 to 2 x n
//Enumerate subcolumns from 1 to n
//Assign adjacent end points to the next non empty subrow
//If A and B are adjacent and B and C are adjacent, then A is also adjacent to C, subsequently C is indirect adjacent to A
//Discard indirect adjacencies
//Because the matrix is symetric, matrix entries above the diagonal are processed only, 
begin							//1
row = 1
column = 2
for each row from 1 to n	//2
  for each column from 2 to n	//3
    for each subrow from 0 to 1	 //4	
      for each subcolumn from 0 to 1	 //5
        if GetAdj(A,n,subrow,subcolumn,row,column) is TRUE	//6
        then
          for each row1 from 1 to n		//7
           for each subrow1 from 0 to 1		//8
            if GetAdj(A,n,subrow1,subcolumn,row1,column) is TRUE	//9
            then
             for each column1 from 1 to n		//10
              if GetAdj(A,n,subrow1,subcolumn,row1,column) is TRUE	//11
              then
               SetAdj(A,n,subrow1,subcolumn,row1,column1,FALSE)
               //SetAdj(A,n,subrow, subcolumn,row, column1,TRUE)
              end if //11
             end for //10 - column1
            end if	 //9
           end for		//8 - subrow1
          end for		 //7 - row1
        end if			 //6
      end for			 //5 - subcolumn
    end for				//4 - subrow
  end for				 //3 - column
end for					//2 - row
return Incidence List
end						 //1
\end{lstlisting}

\subsubsection{CONVERT INCIDENTS MATRIX TO INCIDENCE LIST}

\begin{lstlisting}
procedure IncidentsMatrix2List(A,L,n,m)

//Input: Incidents Matrix A with n Arcs and 2 x n nodes //first row and column are starting //with index = 1
//Output: Incidents List L(arc,source,target)

begin							//1
NewNode = 0
//The two old nodes of arc y are 2 x y and 2 x y +1
//For getting the new nodes the non empty subrows are counted only 
//In the NodeList(OldNode,NewNode,degree), OldNode is 2x(row-1)+subrow+1
//
//Count the degree of each OldNode  
for each row from 1 to n	//2
  SetAdj(A,n,0,0,row,row,FALSE)
  SetAdj(A,n,1,1,row,row,FALSE)
  for each subrow from 0 to 1	//3
    OldNode = 2x(row-1)+subrow+1
    degree = 0
    for each column from 1 to n	 //4
      for each subcolumn from 0 to 1  //5	
        if GetAdj(A,n,subrow,subcolumn,row,column) is TRUE
        then increment degree 
      end for //5
    end for  //4
   end for //3
   if degree > 0   //6
   then
     increment NewNode
     SetNodeList[OldNode,NewNode,degree] 
	 if degree is equal to 1	//7
     then
       SetAdj(A,n,0,0,row,row,TRUE)
       SetAdj(A,n,1,1,row,row,TRUE)
     end if					 //7
   else
     SetNodeList[OldNode,0,0] 
   end if  //6
end if  //2

//Build the Incidents List
OldNode = 0
for each column from 1 to n  //11
  source = 0
  target = 0
  for each row from 1 to n	//12
    for each subrow from 0 to 1  //13
      increment OldNode
      for each subcolumn from 0 to 2  //14
        if GetAdj(A,n,subrow,subcolumn,row,column) is TRUE	//15
        then
          node = GetNodeList(OldNode,NewNode)
          if subcolumn is equal 0
          then source = node
          else target = node 
        end if //15
      end for  //14
    end for    //13
  end for      //12
  if source > 0 and column > 0	//16
  then
    SetIncidentsList(column,source,target)
  end if     //16      
end for		//11
end            //1
\end{lstlisting}

\end{document}